%% file: Paper-sigconf.tex
\documentclass[sigconf]{acmart}

\newcommand{\nz}{New Zealand}
\newcommand{\uoa}{University of Auckland}

\newcommand{\pgcertit}{PGCertInfoTech}

\newcommand{\csoop}{COMPSCI 718}

\newcommand{\csweb}{COMPSCI 719}

\newcommand{\skillshortageNZ}{\url{https://skillshortages.immigration.govt.nz}}

\newcommand{\StackOverflowReport}{\url{https://insights.stackoverflow.com/survey/2020}}
\newcommand{\pgcerttopic}[2]{\textcolor{red}{{#1, #2 hours}}}
\renewcommand{\pgcerttopic}[2]{}
\newcommand{\decision}[1]{\textcolor{green}{#1}}
\renewcommand{\decision}[1]{}

\newcommand{\old}[1]{\textcolor{gray}{Leftover: #1}}
\renewcommand{\old}[1]{}

\newcommand{\qid}[2]{Q#2}

\AtBeginDocument{%
  \providecommand\BibTeX{{%
    \normalfont B\kern-0.5em{\scshape i\kern-0.25em b}\kern-0.8em\TeX}}}

\acmDOI{}

\acmConference{School of Computer Science}
\acmBooktitle{The University of Auckland}
\acmISBN{}

\begin{document}

\title{The Industry Relevance of an IT Transition Programme}

\author{Yu-Cheng Tu, Ewan Tempero, Paramvir Singh, Andrew Meads}
\email{{yu-cheng.tu,e.tempero,p.singh,andrew.meads}@auckland.ac.nz}
\affiliation{%
  \institution{The University of Auckland}
  \city{Auckland}
  \country{New Zealand}
}

\renewcommand{\shortauthors}{Tu, et al.}

\begin{abstract}
  \input{content/abstract}
\end{abstract}





\settopmatter{printacmref=false}
\setcopyright{none}

\maketitle

\section{Introduction}\label{introduction}

\input{content/introduction.tex}



\section{Related Work} \label{related}

\input{content/relatedWork}

\section{Background} \label{background}

\input{content/background.tex}

\section{Research Methodology}\label{methodology}

\input{content/researchMethod}

\section{Results and Analysis}\label{results}

\input{content/results.tex}

\section{Discussion} \label{discussion}

\input{content/discussion.tex}

\section{Conclusions}\label{conclusions}

\input{content/conclusion}

\bibliographystyle{ACM-Reference-Format}
\bibliography{IndustryNeeds}

\end{document}

%% file: content/abstract.tex
There is a shortage of qualified people in the IT industry in the world. To address this shortage, transition programmes are being created that help people change to careers in IT. To provide useful programmes, we need to know if the current curriculum provides value to its graduates. Moreover, as the IT industry undergoes continuous change, we need to regularly review what the industry needs and update any existing programmes as appropriate. In this paper we present the results of a survey of graduates of one such programme, the \pgcertit{} at \uoa{}, with the view to evaluating the currency of the existing programme and to gather data on which to base decisions on updating it. Our conclusion is that our programme is largely useful to graduates, but could be improved with the addition of material on continuous integration, and some adjustment to the time spent on testing, concurrency, and project management. Our results will be useful to any other institutions having, or considering to have, IT transition programmes.

%% file: content/introduction.tex
There is a shortage of qualified people in the IT industry in the world, including 
\nz.\footnote{\skillshortageNZ} To address this shortage, transition programmes have been created that help people change to careers in IT. At the \uoa\/ the \pgcertit{} is one such. The software industry always strives to recruit quality software developers. To provide a useful transition programme, we must understand what these developers need in order to fulfil industry's expectations. Moreover, the IT industry undergoes continuous change, meaning a programme that was acceptable when it was first created may no longer be fit-for-purpose, that is, it no longer provides the necessary training to allow a change in career to what the industry needs today. Consequently we need to regularly review what the industry needs are in order to evaluate our programme. In this paper we present the results of a survey we conducted whose goal was to help us evaluate and, if necessary, update \pgcertit.

The IT industry is very broad, and so no one programme can possibly cover it
all. Thus it was never the intent that the \pgcertit{} could cover everything
that someone working in this industry would need. Instead the goal is to
provide a core set of topics that potential employers of our graduates would
find useful. This means that our primary goal was to evaluate how useful the
topics we teach actually are for our graduates.  To determine how useful
topics are, rather than ask general questions about perceived usefulness, we
instead asked specific questions about actual use.  We surveyed only graduates
from the programme, to ensure that our survey participants were familiar with
the topics we asked about.

The \pgcertit{} programme provides opportunities for people coming from a non-ICT background to gain essential software development skills. The programme is for students who have a bachelor's degree in a non-ICT area and who want to upskill or pursue a career in IT.

The \pgcertit{} programme consists of two courses, \csoop{} and \csweb{}. \csoop{}, \emph{Programming for industry}, focuses on the fundamentals of object-oriented programming using Java. \csweb{}, \emph{Programming with web development}, focuses on developing web applications using various front-end and back-end web technologies. These are intensive courses covering a wide range of topics. We are reviewing the curriculum and our goal is to determine whether the topics in these courses are the right topics we should be teaching now. There are other topics that we might also include, perhaps by reducing the amount of content we currently cover of existing topics. To help us make this decision, we surveyed graduates of the \pgcertit{} programme to determine which, and to what degree, of the topics in the programme our graduates have encountered in their jobs.

In this paper we describe our survey and present our findings from it.  Our general conclusions are that the \pgcertit{} is still acceptable as is, however there are some changes that we could make to improve it, such as changing the amount of material for some topics (either increasing or decreasing), and adding some new topics. We aim to answer the following research questions:
\begin{enumerate}
    \item Does the existing programme provide value to its graduates?
    \item What changes do we propose to our programme based on the needs of industry?
\end{enumerate}

The rest of the paper is organised as follows. In the next section, we present related work in Computer Science (CS) and Software Engineering (SE) education. In Section \ref{background} we summarise the \pgcertit{}. We then present our research methodology in Section \ref{methodology} and results and analysis in Section \ref{results}. We discuss results in Section \ref{discussion}, and then conclude with our findings.

%% file: content/relatedWork.tex
The existing research provides insight into the exploration of the relevance and fitness of ICT programmes, along with their underlying courses, to the industry. Most studies report experiences with designing courses that encompass industry-relevant activities to prepare the students for their various industrial roles \cite{Lundqvist2018,YuCheng2018,Tuzun2018}. These studies analyze students’ feedback to understand the importance of the coursework after delivering the course. However, the decision on the design of such courses and the selection of the course contents involved might be better served by receiving inputs from industry professionals. Hence, the other smaller set of more recent studies follow this approach.

\subsection{Acquiring Industry needs for SE Education}

The existing studies leverage different sources and methods to acquire the software development industry needs for the graduates completing an ICT programme. For instance, \citeauthor{Gurcan&Kose2017}~\cite{Gurcan&Kose2017} used semantic topic analysis on online SE job postings to identify such industry needs. They identified terms representing various roles, responsibilities, high level topics and programming languages, that can potentially guide prospective innovative curriculum design for SE education. Similarly, with an aim to find a balance between the needs of the industrial jobs and the skills acquired in academia, Moreno et al. \cite{Moreno2012} exploited the software engineering knowledge guidelines (SE2004~\cite{Lethbridge2006} and GSwE2009~\cite{Pyster2009}) and the online job postings via Career Space.

However the predominant research methods are surveys (literary or field surveys) and interviews. For instance, \citeauthor{Garousi2019}~\cite{Garousi2019} conducted a systematic literature review on identifying the gaps between SE industry and education. \citeauthor{Gurcan&Kose2017}~\cite{Gurcan&Kose2017} highlighted the skill gaps in the high importance areas of software design, quality, testing and configuration management, among other topics including topics on soft skills. Similar findings and attitudes were reported by other studies that involved either interviews or surveys with industry professionals on SE industrial needs \cite{Watson&Blincoe2017,Liebenberg2015,Kitchenham2005}. Our study does not focus on identifying soft skills required for ICT graduates and hence we did not further explore the relevant research.

All the studies mentioned above involve questions around a wide spectrum of SE and CS related topics, whereas we intend to specifically identify the use in industry of topics that are generally taught or are likely to be added to the future offerings of our \pgcertit{}. Further, we do not directly seek the importance or relevance of each topic for industry rather we aim to derive such knowledge through their actual use in the routine work of industry professionals. 

\subsection{Alumni as participants}

Several studies relied on alumni as participants for their studies. This was particularly the case with studies aiming to verify the industrial relevance of contents of a given SE course. For instance, \citeauthor{Deak&Sindre2017}~\cite{Deak&Sindre2017} conducted two surveys involving their alumni to understand the relevance of software testing related topics. Our study falls in this category as we are looking to study the importance of the contents of our programme to our graduates. 

%% file: content/background.tex
The \pgcertit{}, introduced in 2015, is an intensive and fast-paced programme with the goal of providing industry-focused education to non-ICT students. The programme also provides an opportunity for people to gain a formal qualification in IT. The programme is equivalent to one semester (12 weeks) of full-time study, where students study 40 hours each week. Each lecture is 1--2 hours long. Students must study \csoop{} and \csweb{} to complete the programme. Class starts at 9am with lectures to learn new concepts and finishes at 4pm with a supervised lab to work on practical exercises. The programme also offers part-time study option where students study one course each semester over two semesters.

\csoop{} aims to develop problem-solving skills in the context of software development. The course focuses on object-oriented programming using Java as the primary programming language. The course covers introductory and advanced programming concepts that are typically taught in computer science or software engineering first-year and second-year programming courses. The topics in the course include programming fundamentals, inheritance and polymorphism, exception handling, collections, recursion, concurrency, design patterns, testing and refactoring.

\csweb{} aims to expose students to modern web development technologies. The course covers front-end and back-end web development. The front-end web topics in the course include HTML, CSS, and JavaScript. The back-end web topics include Java Servlets, cookies and sessions, introduction to SQL, and introduction to web security. The course also covers source and version control using Git, web deployment, and remote database servers. The part-time variation of \csweb{} uses NodeJS instead of Java Servlets for the back-end development. 

Table \ref{tab:topics} summarises the topics covered and the lecture hours spent in each course.

\begin{table}[ht]
    \centering
    
    \caption{Topics and hours spent in \pgcertit{}}
    \begin{tabular}{lp{5cm}l}
        \hline
        & \textbf{Topic} &  \textbf{Hours}  \\ \hline
        \textbf{\csoop} & Object-oriented programming (OOP) & 25  \\
        & Unified Modelling Language (UML) & 3 \\
        & Recursion & 2 \\
        & Swing & 3 \\
        & Concurrency and SwingWorker & 6 \\
        & Design Patterns & 6 \\
        & Testing & 3 \\
        & Refactoring & 3 \\
        & Planning using Scrum & 1 \\ \hline
        \textbf{\csweb{}} & HTML & 6  \\ 
        & Cascading Style Sheets (CSS) & 12 \\
        & JavaScript & 9 \\
        & Servlets (or NodeJS) & 3 \\
        & Cookies and Sessions & 3 \\
        & Databases & 6 \\
        & Java Database Connectivity (JDBC) and & \\
        & Data Access Objects (DAOs) & 3 \\
        & Security and Deployment & 3 \\
        & Git & 3 \\ \hline
    \end{tabular}
    \label{tab:topics}
\end{table}

%% file: content/researchMethod.tex
Our general goal is to understand the relevance of \pgcertit{} to industry, that is, those who take our programme find it useful in their new careers. To do so we need to determine what was of value to our graduates. This requires eliciting relevant information from our graduates. The most efficient means to do this is a survey, and so that is what we chose to do. 

When conducting surveys, two important decisions are: what do we ask in the survey, and who do we survey. Given our goals, it seemed most appropriate to survey our graduates, rather than the industry in general. As noted earlier the IT industry is very broad, and the \pgcertit{} was not intended to cover it all. So including participants from industry who were not graduates ran the risk receiving responses not relevant to our programme. Had our goal been to expand the \pgcertit{} this would have been the better option. There would also have been the issue of getting a representative sample as contacting relevant participants would not be easy. As we had contacts for all of our graduates this seemed the best population to survey.
We could be sure that this group had been exposed to the topics we were interested in, and provide insights to topics that might be the most relevant to the industry.

Our goal was to determine whether or not the topics taught in the programme were useful. Accordingly it is appropriate that a significant number of the questions we ask relate to the courses as they have been taught so far. 

To develop the survey, we initially considered the topics which are taught, to some degree, in \pgcertit{}. However, as technology changes, we also considered some topics not currently taught that are currently popular in industry\footnote{\StackOverflowReport}\cite{Garousi2019-2}. For large topics, such as Object-Oriented Programming, appropriate subtopics were identified (e.g. use of inheritance, polymorphism). For Object-Oriented Programming, we also asked questions to get a sense of the degree to which they used relevant concepts (see Section \ref{results.oop}). The list of topics covered in our survey can be seen in Table \ref{tab:surveyTopics}.

\begin{table}[ht]
    \centering
    \caption{List of topics in the survey}
    \begin{tabular}{ll}
        \hline
        \textbf{Topics} & \textbf{Subtopics} \\ \hline
        Concurrency & Asynchronous execution \\ 
        & Explicit use of threads \\ \hline
        Continuous integration & \\ \hline
        Databases & Relational databases \\
        & NoSQL \\ \hline
        Design patterns & \\
        Documentation & \\
        Frameworks & \\  \hline
        Graphical User Interface (GUI) & \\  \hline
        OOP & Use of inheritance \\
        & Encapuslation \\
        & Use of polymorphishm \\  \hline
        Project management & \\  \hline
        Recursion & \\  \hline
        Refactoring & \\  \hline
        Security & \\  \hline
        Testing & Unit testing \\
        & Automated testing \\
        & Test-driven development \\  \hline
        Version control & \\  \hline
    \end{tabular}
    \label{tab:surveyTopics}
\end{table}

How we asked our questions was important. We could have asked participants’ perceptions of whether or not a topic is important, but this would have introduced a degree of subjectivity in participants' responses. Instead  our goal is to ascertain the extent to which our participants use that topic in their day-to-day employment. Doing so meant that participants' responses were based on concrete experience rather than opinion. 

To this end, the majority of survey questions are of the form “In your employment, have you used [topic / subtopic] within [timeframe]?” or “Have you used [topic / subtopic] in your employment?”. Where further categorization or clarification is required, follow-up questions were asked. For example, after determining that a participant used version control, a follow-up question would be asked to elicit the specific version control tool used (e.g. git, svn). Open-ended free-text-entry questions were also added for each topic of the form “Please provide any further comments on your use of [topic / subtopic]”.

We also gather demographic data to capture the experience our participants had both since completing the programme and any previous experience they had. This was necessary to interpret the responses. The survey contains 52 questions in total.

Prior to releasing the survey, a small pilot was conducted where two participants in the target demographic filled out the survey in addition to providing feedback on the survey’s length and content. The feedback indicated that no changes needed to be made.

We have received ethics approval from the University to conduct the survey in January 2020. We distributed the survey via an alumni mailing list maintained by the \uoa. There were 148 graduates at the time of the distribution of the survey. The distribution was performed by the institution administration. The authors were not involved in this process.

%% file: content/results.tex
As discussed in Section \ref{introduction}, we are looking for data to help us
make decisions about possible changes to the courses. In this section we
present just the raw data together with points of interest.  We will discuss
how we might use this data in making decisions in Section \ref{discussion}.

The overall results are shown in Table \ref{table:results}. The QID labels
indicate the topic and will be used in the text below. Due to space
constraints we report just the number of participants who answered positively
to the questions we asked. While 48 started our survey, only 27 completed
it. Our results are presented for those 27.

\begin{table*}
  \caption{Positive responses to questions (from 27 participants). QID explained in text. \label{table:results}}
  \begin{tabular}{|l||rr|rr|rrrr|rr|r|rr||} \hline
    \textbf{QID}
    & \qid{Q14}{cc1} & \qid{Q16}{cc2}
    & \qid{Q34}{ci1} & \qid{Q35}{ci2}
    & \qid{Q29}{db1} & \qid{Q30}{db2} & \qid{Q31}{db3} & \qid{Q32}{db4}
    & \qid{Q18}{dp1} & \qid{Q19}{dp2}
    & \qid{Q10}{doc}
    & \qid{Q27}{fw1} & \qid{Q28}{fw2}
    \\ 
    \textbf{Yes}
    & 20 & 17
    & 14 & 10
    & 20 & 24 & 13 & 4
    & 19 & 10
    & 12
    & 20 & 17
    \\
    \hline
  \end{tabular}
  \begin{tabular}{|l||r|rrrr|r|r|r|r|rrrr|r||} \hline
    \textbf{QID}
    & \qid{Q17}{gui}
    & \qid{Q40}{oop1} & \qid{Q7}{oop2} & \qid{Q5}{oop3} & \qid{Q6}{oop4}
    & \qid{Q36}{pm}
    & \qid{Q13}{rec}
    & \qid{Q24}{rf}
    & \qid{Q33}{sec}
    & \qid{Q20}{t1} & \qid{Q21}{t2} & \qid{Q22}{t3} & \qid{Q23}{t4}
    & \qid{Q25}{vc} 
    \\ 
    \textbf{Yes}
    & 16
    & 23  & 11 & 15 & 9
    & 24 
    & 19
    & 24
    & 18
    & 8 & 9 & 13 & 14
    & 24
    \\
    \hline
  \end{tabular}
\end{table*}

\subsection{Demographics}

Of the 27 participants, 18 included ``Developer'' as their role (some listed
other roles such as ``Architect''), 3 gave ``Consultant'' as their role, 2
gave ``Business Analyst'', and 4 gave other roles.

We asked what experience participants had in IT related areas. There were 13
with less than 1 year, 8 with 1-3 years, 4 with 3-6 years, and 2 with 6-10
years. As the \pgcertit{} was first delivered in 2015, the two participants
with 6-10 years of experience were likely to already have some work experience
prior entering our programme. The programme to provided a pathway for them to gain formal qualifications in IT.

\subsection{Concurrency}

\pgcerttopic{718}{6} \decision{reduce, remove threading, keep async stuff}

We asked about developing code involving concurrency. We divided this into two
subtopics --- development of code involving concurrency through asynchronous
execution, where the concurrency is implicit (\qid{Q14}{cc1}), and development
of code where the concurrency is explicit through the use of multiple threads
of control (\qid{Q16}{cc2}). We found the use of asynchronous concurrency is
noticeably more common (20), although use of explicit threads was not uncommon (17).

\subsection{Continuous Integration} \decision{add}

We asked about continuous integration and related topics, specifically
regarding use of automated testing during builds (\qid{Q22}{t3}), creation of
tests other than unit tests (\qid{Q23}{t4}), use of continuous integration in
the organisation (\qid{Q34}{ci1}), and use of DevOps (\qid{Q35}{ci2}).

Continuous integration is currently not taught in our courses. About half of participants reported some exposure to this topic. These results suggest that we should consider at least exposing the students to the relevant concepts.

\subsection{Databases} \pgcerttopic{719}{9} \decision{no change}

We asked about development involving databases, including writing code to
access databases (\qid{Q29}{db1}), use of SQL (\qid{Q30}{db2}), designing
relational databases (\qid{Q31}{db3}), and use of NoSQL (\qid{Q32}{db4}).
Most had written code (20) and used SQL (24).  Of particular note is that
\qid{Q29}{db1} asked about writing code accessing a database \emph{in the last
  month}, meaning the relatively high positive response probably under reports
the amount of such programming done.  Our results indicate that exposure to
relational database programming is important.

With \qid{Q29}{db1}, we also asked which database management systems
participants had use. The common choices were 13 for MS SQL, 10 Mysql, and 5
postgres (participants could report more than 1 system).

We were a little surprised that there was not more use of NoSQL, given the
prominence it has received\footnote{https://www.alliedmarketresearch.com/NoSQL-market}.

\subsection{Design Patterns}

\pgcerttopic{718}{6} \decision{reduce, focus on GUI}

We asked whether participants had encountered use of design patterns
(\qid{Q18}{dp1}) and if they had applied them themselves (\qid{Q19}{dp2}).
The majority had seen them in their projects, and of those 10 reported
applying them. The main patterns mentioned were: MVC (7), Singleton (6),
Factory (5), Observer (5) and Template Method (4). 

\subsection{Documentation}

\pgcerttopic{718}{3} \decision{reduce UML, add general inline commenting}

We were interested generally in what documentation, if any, participants dealt
with. We asked whether they had been exposed to diagrams (\qid{Q10}{doc}).
Only 12 used diagrams, of which 4 had used UML, 2 had used ER diagrams, and 6
had used both. All but one participant (an Intern) had used some form of
documentation. Various forms of documentation were reported, including inline
code comments (20), use of Wiki (14), method comments (13), and JavaDoc
(6). All but 4 reported using multiple forms of documentation, with 3 of those
4 using Wikis.

\subsection{Frameworks}

\pgcerttopic{719}{6} \decision{no change}

We asked what kinds of frameworks participants had seen used for front-end
development (\qid{Q27}{fw1}) and back-end development (\qid{Q28}{fw2}). Of
those who had seen front-end frameworks (20), there was a fairly even spread
between Angular (7), Bootstrap (9), JQuery (8), and React (8), with some
others mentioned. Eleven participants had used more than one such framework.

For back-end frameworks 17 reported using them and there was a wider variety
reported, including .NET Core (8), Flask (4), and Express (3). Only 5
participants reporting using more than one framework. All 17 also reported using
front-end frameworks.

\subsection{GUI}

\pgcerttopic{718}{6} \decision{no change}

We asked whether participants had written code that responds to events
(\qid{Q17}{gui}). There were 16 who reported having done so, and of these 8
had experience applying either the MVC or the Observer pattern (see
\qid{Q19}{dp2}).

\subsection{Object-Oriented Programming} \label{results.oop}

\pgcerttopic{718}{30} \decision{no change}

We asked participants whether they had worked with OOP (\qid{Q40}{oop1}),
whether they had seen poorly encapsulated code (\qid{Q7}{oop2}), whether they
had worked with inheritance (\qid{Q5}{oop3}), and whether they had used
polymorphism (\qid{Q6}{oop4}).

We found almost all of our participants (23) reported that they have written
or modified code written in object-oriented languages in their employment since
graduation. This is not surprising as object-oriented programming is the focus
of the programme, and so we would expect most graduates to find positions
consistent with having this knowledge. Somewhat surprising was how few
reported using polymorphism.

We also asked what languages participants used.  The most popular languages
were JavaScript with classes (13), Python (12), Java (11), C\# (9), and C++
(3). More than half of our participants (17) reported more than one programming language
used at their work.

Of those who had not written any OOP code, one reported their role as an
Intern, one as a Spatial analyst, and one as a Business Analyst. The 4th
participant listed their role as a developer, but had only worked with Python
and HTML.

One concern with the wording of question \qid{Q40}{oop1} was that participants
who had only cursory experience with OOP in their employment would answer
positively, which might lead to mis-leading results. To address this, we asked
a set of questions to gain more insight as to the degree of their
experience. For those answering positively to \qid{Q40}{oop1}, we asked
whether it was only reading code (2 of 23 answered positively), and for those
who wrote code, whether they had created classes, abstract classes, or
interfaces in the last month (13/21). As we specified only classes and
interfaces created in the last month, it is likely that more participants do
so during their employment. Our question on inheritance (\qid{Q5}{oop3}) did
not limit to use in the last month, and three who answered negatively to
\qid{Q40}{oop1} answered positively to this question, confirming this belief.

\subsection{Project Management}

\pgcerttopic{719}{1} \decision{increase from one hour, more than one technique}

We asked what project management methodologies participants were exposed to
(\qid{Q36}{pm}). Most were exposed to an agile methodology, with Scrum the most common
(19) followed by Kanban (9). Interestingly 5 reported using Waterfall, and 12
used more than one methodology. We also asked what project management tools participants
used. Common tools were Jira (16), Microsoft Teams (14), Slack (11), Trello
(9), with 19 reporting using more than one tool.

\old{Participants also commented on the use of different project management
methodologies for different projects.}

\subsection{Recursion}

\pgcerttopic{718}{3} \decision{no change}

We asked participants about their experience with recursion in their work
(\qid{Q13}{rec}).  The majority of participants (19) reported either using it
(12) or encountering it in code (7). In addition 3 saw situations where
recursion was not used but would have been appropriate.

\subsection{Refactoring}

\pgcerttopic{718}{3} \decision{no change}

We asked whether participants refactored their code (\qid{Q24}{rf}). The
majority (24) reported they did. We also asked their reasons for doing so. As other surveys had indicated many in industry faced barriers to doing refactoring \cite{TemperoGorschekAngelis:CACM:2017} we
expected the majority would be due to organisational requirements (e.g. a consequence of the use of
code review), but only 6 indicated this reason.

\subsection{Security} \pgcerttopic{719}{3} \decision{no change}

We asked if participants had to deal with security issues when writing code
for their organisation (\qid{Q33}{sec}). About two thirds indicated that they
had.

\subsection{Testing}\label{topic.testing}

\pgcerttopic{718}{3} \decision{reduce tdd, focus on testing for continuous integration}

We asked if test-driven development was used in participants' organisation
(\qid{Q20}{t1}), whether they had written unit tests (\qid{Q21}{t2}), if
automated testing was part of their organisation's build process
(\qid{Q22}{t3}), and if they had written any other kinds of tests
(\qid{Q23}{t4}). Not more than a third wrote unit tests however nearly half
wrote other kinds of tests. For one third of participants, their organisation
used test-driven development, and half were in organisations that used
automated testing in their builds.

\subsection{Version Control}

\pgcerttopic{719}{6} \decision{no change}

We asked what version control systems participants used, if any
(\qid{Q25}{vc}).  Most reported using something. Git was the most common
choice (21) and all other systems mentioned only by one participant each.

We also asked how participants accessed the version control systems.  The most
common was IDE (17) followed by from the command-line (12), with 10 using multiple methods.

%% file: content/discussion.tex
The results suggest that by and large the topics we already have are fit for purpose (RQ1). The results suggest no changes to databases, frameworks, GUI, object-oriented programming, recursion, refactoring, security, and version control. Our results indicate that exposure to these topics are important in industry.

The results give us confidence to teach object-oriented programming in \csoop{} as most of our participants have reported working with code written in object-oriented languages. We noticed only few reported using polymorphism which seemed to be inconsistent with the number of participants reported working with object-oriented languages. Given many participants have not much experience, we suspect that they probably have been using polymorphism at work without knowing, suggesting we may need to discuss this more in the course.

For RQ2, the topics we should consider adjusting are concurrency, design patterns, and project management. The results suggest the use of asynchronous concurrency is more common than use of explicit threads. Currently, \csoop{} covers concurrency with the use of explicit threads and SwingWorker. Based on the results, we think it would be appropriate to reduce the number of hours on explicit threads in \csoop{}, and introduce asynchronous concurrency in \csweb{} as the topic is more applicable to web development. 

The other topic to adjust is the design patterns taught in \csoop{}. The course currently covers adapter, composite, observer and template method patterns. The results show that while these design patterns are relevant to the industry, they were not the most frequently mentioned. Two people specifically mentioned MVC as important:
\begin{quote}
``MVC is the most important that everybody should know. The rest don't need to be explicitly taught.''
\end{quote}
From our results we conclude that introducing patterns such as MVC and Singleton would seem more appropriate for \csoop{}.  

Currently, we spend one hour on planning using Scrum in \csoop{}. Based on the results, we think more time should be spent on project management in \csoop{} as well as in \csweb{}. As most participants were exposed to at least one agile methodology, other agile methodologies such as Kanban should be introduced.

We also think testing should be adjusted in \csoop{}. To our surprise, not more than a third wrote unit tests and only one third of participants used test-driven development in their organisations. The results do not indicate unit testing or test-driven development being used as much as we expected. One participant, who has 6 – 10 years of experience as a developer, explained why they did not use test-driven development:

\begin{quote}
    ``Really depends on the pace of the company - a SME with about 20 devs moves too fast for TDD to be useful, as we rely on failing quickly to find issues that would otherwise take a lot of time to find (in our opinion).''
\end{quote}

Some of our participants mentioned that they usually have a separate team writing unit tests or their roles do not require writing unit tests. They write unit tests only if they have to do so. Several participants commented on testing at their workplace:
\begin{quote}
    ``Currently they employ separate QAs who write tests concurrently as the developers write the code. However I think they are considering moving towards test-driven development.''
\end{quote}
\begin{quote}
    “Usually the unit tests and automated tests are written and run by a QA. We are expected to have `dev-tested' it ourselves before submitting to the QA. However I did spend one week writing unit tests.”
\end{quote}
\begin{quote}
    “I very rarely unit test myself because it's not possible for the infrastructure I work on, but other devs do and we require it as part of our application build process...”
\end{quote}

Half of our participants also indicated the use of automated testing at work, which is an important element of continuous integration. In addition, the results reveal a number of participants have seen or used DevOps at work. Based on the results, we need to rethink what we teach in testing in \csoop{} and consider at least exposing the concept of continuous integration and DevOps in either \csoop{} or \csweb{}. 

Documentation is a topic that we feel needs more investigation. The results do not give us a strong indication of the types of documentation that developers use at work. All but one participant had used some form of documentation, including inline comments. One participant made a comment on documentation:
\begin{quote}
    “There’s never enough documentation. Generally documentation is stored in sharepoint, wiki or on a shared harddrive. Inline comments are a must for most jobs. GIT commits also a must”
\end{quote}

It is worth commenting on our decision to ask about recursion. On the one hand, this is a fundamental concept in computer science and so would seem reasonable to teach, on the other hand there are a number of fundamental computer science topics that are not taught in the \pgcertit{} so we wondered whether there was justification for including recursion and not these others topics. There was speculation within our team that it was not actually used in industry sufficiently to justify its inclusion and so we included a specific question on it to test this. The results clearly show that this is a concept worth covering in the programme. In fact, one participant implicitly indicated the importance industry attaches to this topic:
\begin{quote}
    ``Also had an interview question on recursion''
\end{quote}

The results of our survey are helpful for us to evaluate our existing programme. In some cases the results might support multiple decisions (e.g. either increasing or decreasing content), but we can now make those decisions informed to some degree by what we know is happening in industry. 

As with any survey-based research there are caveats. The most obvious question, and one common to any survey, is whether our sample was representative of the population. Because we limited our population, we were then able to contact almost everyone (148 in total). Of those contacted 48 started and 27 completed it, so a response rate of 18\%. Whether this is sufficient is a matter of debate. Singer
et al. report a consistent response rate of 5\% in software engineering
studies \cite{Singer2008} so ours compares well with that. 

However what is most important is whether the conclusions drawn from the responses could be invalidated by the non-responses. A key point is that our data is a \emph{lower bound} on the usage of the different topics. If one person reports using a concept then we know at least one person in our population uses the
concept, and it could easily be more. What we conclude from this, in particular whether we think this one usage is sufficient evidence to justify its inclusion in the \pgcertit{}. is then dependent on our specific context. Other programmes, on seeing this evidence, may draw a different conclusion because they have a different context. 

We only surveyed graduates of our programme. These graduates are likely to have sought positions that match the skills they have acquired from our programme, and so perhaps the responses to the topics in our survey are unsurprising. However what our results show is that graduates of our programme have, by and large, used the content provided in our programme, which gives us confidence it does provide value.

Our programme, and most of our graduates, exists in the \nz{} context. Nevertheless the original design of our programme was informed by international experience, and other research (Section \ref{related}) suggests the differences between countries is not great. Therefore we believe our results will be useful for others considering developing or revising similar programmes.

Our survey was based primarily on the existing topics in the \pgcertit{} with
only a few other topics that we thought we should consider. There are certainly other topics we might have included. However any choices made for any curriculum is a trade-off of topics and available resources (including time). The last question of our survey asked whether there were topics used in their work that we had not asked about. Five answered this question, mentioning people skills, use of frameworks (2), performance issues, and use of the command line.

The comment about command line was particularly interesting. It was:
\begin{quote}
    ``I reckon it's really important to know how to navigate commandline comfortably, as a huge majority of development tools are commandline only''
\end{quote}
This opinion was supported by the responses to the question on version control (\qid{Q25}{vc}), where a number of participants accessed their version control through the use of the command line. We currently do not specifically teach use of the command line. Based on these results we will consider introducing command line concepts to the programme.

%% file: content/conclusion.tex
In this paper we present the results of a survey of graduates of an IT transition programme 
 with the view to evaluating its currency and to gather data on which to base decisions on updating it. Our conclusion is that our programme, the \pgcertit{} at \uoa{} is largely relevant to our graduates working in the industry, but could be improved with adjustments in \csoop{} to the coverage of concurrency, design patterns and testing, while \csweb{} could introduce new topics such as asynchronous concurrency, continuous integration and project management. 

For other institutions considering introduction a transition programme such as ours, they would likely come up with a similar set of topics to what we have. However, we hope that the data we have given will help provide more confidence with their decisions. For institutions who already have such programmes, we hope our data will help with their evaluation.